\RequirePackage{lineno}

\documentclass[12pt,preprint]{emulateapj}

\newcommand{\Mearth}{$M_\oplus$}

\usepackage{amsmath}

\slugcomment{}

\shorttitle{On the volatile enrichments and heavy element content in HD 189733b}

\shortauthors{Mousis et al.}

\begin{document}

\title{On the volatile enrichments and heavy element content in HD 189733b}

\author{O.~Mousis,$^{1}$ J.~I.~Lunine,$^{2}$ J.-M.~Petit,$^{1}$ K.~ Zahnle,$^{3}$ L.~Biennier,$^{4, 5}$ S.~Picaud,$^{1}$ T.~V.~Johnson,$^{6}$ J. B. A.~Mitchell,$^{4, 5}$ V.~Boudon,$^{7}$ D.~Cordier,$^{4, 5, 8}$ M.~Devel,$^{9}$ R.~Georges,$^{4,5}$ C.~Griffith,$^{10}$ N.~Iro,$^{11}$ M.~S.~Marley,$^{3}$ and U.~Marboeuf$^{12}$}

\email{olivier.mousis@obs-besancon.fr}

\address{$^{1}$ Institut UTINAM, CNRS-UMR 6213, Observatoire des Sciences de l'Univers THETA, Universit\'e de Franche-Comt\'e, BP 1615, 25010 Besan\c{c}on Cedex, France \newline
$^{2}$ Dipartimento di Fisica, Universit{\`a} degli Studi di Roma ``Tor Vergata'', Roma, Italy \newline 
$^{3}$ NASA Ames Research Center, Moffett Field, CA 94035, USA \newline
$^{4}$ Universit\'e europ\'eenne de Bretagne, Rennes, France \newline
$^{5}$ Institut de Physique de Rennes, CNRS, UMR 6251, Universit{\'e} de Rennes 1, Campus de Beaulieu, 35042 Rennes, France \newline
$^{6}$ Jet Propulsion Laboratory, California Institute of Technology, Pasadena, CA 91109, USA \newline
$^{7}$ Laboratoire Interdisciplinaire Carnot de Bourgogne, UMR 5209 CNRS-Universit{\'e} de Bourgogne, 9 Avenue Alain Savary, BP 47870, F-21078 Dijon Cedex France \newline
$^{8}$ Ecole Nationale Sup{\'e}rieure de Chimie de Rennes, CNRS, UMR 6226, Avenue du G\'en\' eral Leclerc, CS 50837, 35708 Rennes Cedex 7, France \newline
$^{9}$ FEMTO-ST, CNRS, UFC, UTBM, ENSMM, Besan\c con, France \newline
$^{10}$ Lunar and Planetary Laboratory, University of Arizona, Tucson, AZ, USA \newline
$^{11}$ NASA/Goddard Space Flight Center, Greenbelt, MD, USA \newline
$^{12}$ Laboratoire de Plan{\'e}tologie de Grenoble, Universit{\'e} Joseph Fourier, CNRS INSU, France}

\begin{abstract}
Favored theories of giant planet formation center around two main paradigms, namely the core accretion model and the gravitational instability model. These two formation scenarios support the hypothesis that the giant planet metallicities should be higher or equal to that of the parent star. Meanwhile, spectra of the transiting hot Jupiter HD189733b suggest that carbon and oxygen abundances range from depleted to enriched with respect to the star. Here, using a model describing the formation sequence and composition of planetesimals in the protoplanetary disk, we determine the range of volatile abundances in the envelope of HD189733b that is consistent with the 20--80 \Mearth~of heavy elements estimated to be present in the planet's envelope. We then compare the inferred carbon and oxygen abundances to those retrieved from spectroscopy and we find a range of supersolar values that directly fit both spectra and internal structure models. In some cases, we find that the apparent contradiction between the subsolar elemental abundances and the {mass of heavy elements predicted in HD189733b by internal structure models} can be explained by the presence of large amounts of carbon molecules in the form of polycyclic aromatic hydrocarbons and soots in the upper layers of the envelope, as suggested by recent photochemical models. A diagnostic test that would confirm the presence of these compounds in the envelope is the detection of acetylene. Several alternative hypotheses that could also explain the subsolar metallicity of HD189733b are formulated, among which the possibility of differential settling in its envelope, the presence of a larger core that did not erode with time, a mass of heavy elements lower than the one predicted by interior models, a heavy element budget resulting from the accretion of volatile-poor planetesimals in specific circumstances, or the combination of all these mechanisms.

\end{abstract}


\section{Introduction}

Favored theories of giant planet formation center around two main paradigms, namely the core accretion model (Safronov 1969; Goldreich \& Ward 1973; Pollack et al. 1996) and the gravitational instability model (Cameron 1978; Boss 1997). In the frame of the core accretion model, a solid core forms from the accretion of planetesimals and becomes massive enough ($\sim$5 to 10 \Mearth) to initiate runaway gravitational infall of a large gaseous envelope in which gas-coupled solids continue their accretion (Alibert et al. 2005a; Hubickyj et al. 2005; Mordasini et al. 2009). This model provides the large amount of heavy elements necessary to explain the supersolar metallicities observed in Jupiter and Saturn via the accretion of planetesimals in their envelopes (Gautier et al. 2001; Saumon \& Guillot 2004; Alibert et al. 2005a; Mousis et al. 2006, 2009a). In the frame of the gravitational instability model, gas giant protoplanets form rapidly through a gravitational instability of the gaseous portion of the disk and then more slowly contract to planetary densities (Boss 1997, 2005). In this scenario, due to the limited efficiency of planetesimals accretion during the planet formation, its metallicity should be slightly higher or equal to that of the parent star (Helled \& Bodenheimer 2010).

A puzzling feature of the transiting hot Jupiter HD 189733b (M = 1.15 $\pm$ 0.04 M$_J$) orbiting a K2V stellar primary at the distance of 0.03 AU (Bouchy et al. 2005) is its metallicity, {whose estimates have been found to range} between subsolar to supersolar (see Fig. 1) from determinations of carbon and oxygen atmospheric abundances . {Because the metallicity of the parent star is solar ([Fe/H] = -0.03 $\pm$ 0.04; Bouchy et al. 2005), we still compare the metallicity of HD189733b to this value in the following.} Several sources of data were used by two different groups in order to retrieve these volatile abundances. A first set of data was collected by Swain et al. 2009 (hereafter S09) from the dayside spectrum of HD 189733b with HST NICMOS spectrophotometry in the 1.5--2.5 $\mu$m range, leading them to find subsolar carbon and oxygen abundances. In contrast, from the same set of data {and with their own model}, Madhusudhan \& Seager 2009 (hereafter MS09) found these spectra consistent with supersolar C and O abundances. However, MS09 also found that these species could be in subsolar abundances from spectra of the planet's atmosphere during secondary eclipses with Spitzer broadband photometry. This huge variation of elemental abundances derived by the two groups is due to a wide variety of atmospheric pressure-temperature profiles that are found consistent with the planet's spectra (S09; MS09).

\begin{figure}
\begin{center}
\resizebox{\hsize}{!}{\includegraphics[angle=0]{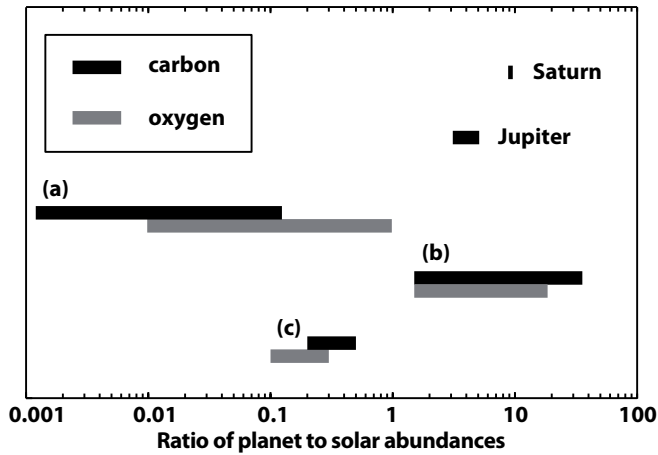}} 
\caption{Carbon and oxygen abundances (relative to hydrogen) in HD 189733b compared with solar values. Elemental abundances derive from the analysis by MS09 (bars a) of the planet spectra acquired with Spitzer broadband photometry during secondary eclipses, and from the interpretation by MS09 (bars b) and S09 (bars c) of the same spectra obtained during dayside observations by HST NICMOS spectrophotometry in the 1.5--2.5 $\mu$m range. Carbon abundances measured in Jupiter and Saturn are shown for comparison (values taken from Mousis et al. 2009a). The oxygen abundance is still unconstrained in these two planets.}
\end{center}
\label{cool}
\end{figure}

Although consensus has not yet been reached on the metallicity of HD 189733b, the possibility that this -- or potentially other -- hot Jupiters\footnote{O and C minimum elemental abundances in the transiting hot Jupiter HD209458b have also been found $\sim$0.3 and 0.7 times solar, respectively (MS09).} have subsolar metallicity raises the challenging theoretical question of whether and how a giant planet may achieve subsolar metallicity during its formation. In order to account for this discrepancy, Mousis et al. (2009b) have proposed that gravitational settling, due to strong irradiation, could lower the carbon and oxygen abundances in the upper layers of HD189733b's atmosphere. However, this possibility has not been yet tested by models detailing the envelope's evolution under the influence of irradiation and might be contradicted by the {possible increase of the atmospheric opacities that} is required to explain the large radii of irradiated planets (Burrows et al. 2007).

In this work, we determine the range of volatile abundances in the envelope of HD189733b that matches the 20--80 \Mearth~mass range of heavy elements predicted by the interior models of Guillot (2008). The latter rejected the models predicting masses of heavy elements lower than 20 \Mearth~in HD189733b on the basis that they are not able to explain in a consistent manner the observed radius measurements of all known transiting giant exoplanets. We also use a model describing the formation sequence of planetesimals in the protoplanetary disk and the composition of the incorporated ices, assuming that the distribution of heavy elements is homogeneous within the planet's envelope. We then compare the inferred carbon and oxygen abundances to those retrieved from spectroscopy and we infer the range of supersolar values that can directly fit both spectra and internal structure models. We also investigate the role that can be played by carbon molecules in the form of polycyclic aromatic hydrocarbons (PAHs) and soots possibly present in the upper layers of the envelope (Marley et al. 2009; Zahnle et al. 2010) in the apparent contradiction between the subsolar elemental abundances and the important mass of heavy elements predicted in HD189733b. We finally discuss the alternative {mechanisms} that could explain this possible discrepancy.


\section{Delivery of heavy elements to proto-HD189733b}

\subsection{Formation conditions of HD189733b}

Irrespective of the details of their formation, close-in giant planets are thought to have originated in the cold outer region of protoplanetary disks and migrated inwards until they stopped at closer orbital radii to the star (Goldreich \& Tremaine 1980; Lin et al. 1996; Fogg \& Nelson 2005, 2007; Mandell et al. 2007). In this context, it has been proposed that core accretion is a method by which planets may form at small distances to the star ($\sim$10 AU) whilst gravitational instability may be the mechanism by which planets may form at much larger distances ($\ge$100 AU) (Boley 2009; Meru \& Bate 2010). Because a bimodal distribution of gas giant planet semi-major axes should remain present after scattering and planet-disk interaction (Boley 2009), implying that giant planets formed by core accretion should migrate closer to the star than those formed by gravitational instability, we follow here the core accretion model to describe the formation of HD189733b. In this scenario, building blocks accreted by proto-HD 189733b may have formed all along its radial migration pathway in the protoplanetary disk. However, in this work, we assume that only the planetesimals produced beyond the snow line, i.e. those possessing a significant fraction of volatiles, materially affected the observed O and C abundances due to their vaporization when they entered the envelope of the planet. This hypothesis is supported by the work of Guillot \& Gladman (2000) who showed that planetesimals delivered to a planet owning a mass similar or greater than that of Jupiter are rather ejected than accreted. This mechanism should then prevent further noticeable accretion of solids by the planet during its migration below the snow line.

\subsection{Composition of icy planetesimals}

In order to compare the amount of heavy elements inferred from measurements of elemental abundances in HD189733b with theoretical determinations of the planet metal content, we have used a model describing the formation sequence and the composition of the different ices formed beyond the snow line of the protoplanetary disk from which HD189733b was formed (Mousis et al. 2009b). This model has already been used to interpret the observed volatile enrichments in the atmospheres of Jupiter and Saturn in a way consistent with the heavy element content predicted by interior models (Mousis et al. 2009a). It is based on a predefined initial gas phase composition in which all elements are in solar abundance (Lodders 2003; see Table 1), and describes the process by which volatiles are trapped in icy planetesimals formed in the protoplanetary disk. {Oxygen, carbon, nitrogen and sulfur are postulated to exist only in the form of H$_2$O, CO, CO$_2$, CH$_3$OH, CH$_4$, N$_2$, NH$_3$ and H$_2$S with CO/CO$_2$/CH$_3$OH/CH$_4$~=~70/10/2/1, N$_2$/NH$_3$ = 1/1 and H$_2$S/H$_2$ = 0.5 $\times$ (S/H$_2$)$_{\odot}$ in the gas phase of the disk (values taken from Mousis et al. 2009a who investigated the composition of ices formed in the protosolar nebula and accreted by Jupiter and Saturn).} Once the abundances of these molecules have been fixed, the remaining O gives the abundance of H$_2$O.

\begin{table}
\caption[]{Gas phase abundances in the solar nebula.}
\begin{center}
\begin{tabular}{lclc}
\hline
\hline
\noalign{\smallskip}
Species X &  (X/H$_2$)  & species X  & (X/H$_2$) \\	
\noalign{\smallskip}
\hline
\noalign{\smallskip}
O		& $1.16 \times 10^{-3}$		& N$_2$		&  $5.33 \times 10^{-5}$ \\
C		& $5.82 \times 10^{-4}$ 		& NH$_3$   	&  $5.33 \times 10^{-5}$ \\
N   		& $1.60 \times 10^{-4}$    		& CO      		& $4.91 \times 10^{-4}$ \\
S        	& $3.66 \times 10^{-5}$		& CO$_2$  	& $7.01 \times 10^{-5}$ \\
P		& $6.88\times 10^{-7}$		& CH$_3$OH  	& $1.40 \times 10^{-5}$ \\ 
Ar       	& $8.43 \times 10^{-6}$		& CH$_4$  	& $7.01 \times 10^{-6}$ \\ 	
Kr       	& $4.54 \times 10^{-9}$		& H$_2$S    	& $1.83 \times 10^{-5}$ \\
Xe       	& $4.44 \times 10^{-10}$		& PH$_3$    	& $6.88 \times 10^{-7}$ \\
H$_2$O  	&  $5.15 \times 10^{-4}$		&							     \\
\hline
\end{tabular}
\end{center}
Elemental abundances derive from Lodders (2003). Molecular abundances result from the distribution of elements described in the text.
\label{lodders}
\end{table}

\begin{figure}
\begin{center}
\resizebox{\hsize}{!}{\includegraphics[angle=0]{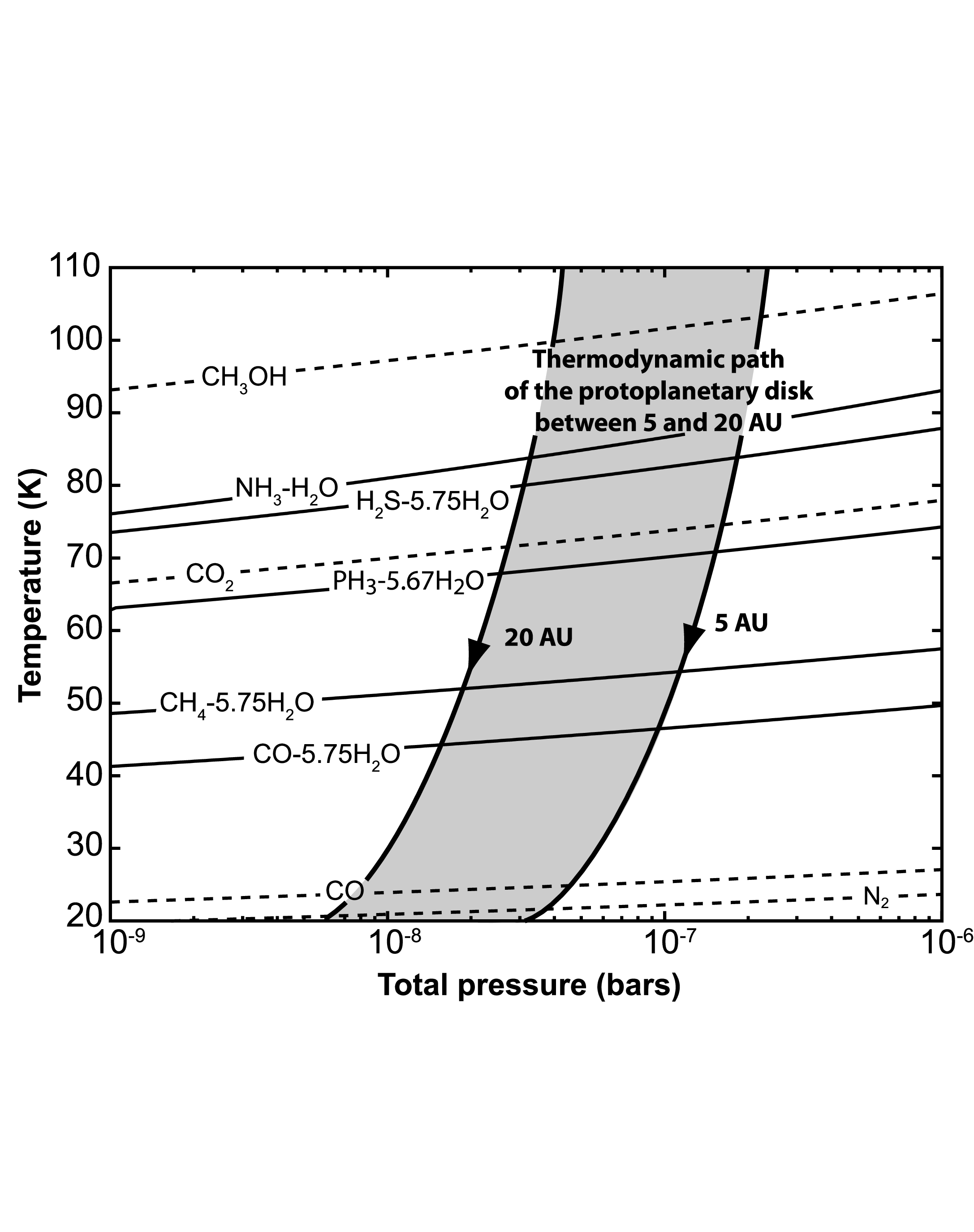}} 
\caption{Equilibrium curves of hydrate (NH$_3$-H$_2$O), clathrates (X-5.75H$_2$O or X-5.67H$_2$O) (solid lines), pure condensates (dotted lines) and the ensemble of thermodynamic paths (grey area) ranging between 5 and 20 AU in the pressure-temperature space of the disk, assuming full efficiency of clathration. Species remain in the gas phase above the equilibrium curves. Below, they are trapped as clathrates or simply condense.}
\end{center}
\label{cool}
\end{figure}

The process of volatile trapping in planetesimals, illustrated in Fig. 2, is calculated using the stability curves of hydrates, clathrates and pure condensates, and the ensemble of thermodynamic paths detailing the evolution of temperature and pressure in the 5--20 AU range of the protoplanetary disk. We refer the reader to the works of Papaloizou \& Terquem (1999) and Alibert et al. (2005b) for a full description of the turbulent model of accretion disk used here. This model {postulates} that viscous heating is the predominant heating source, {assuming} that the outer parts of the disk are protected from solar irradiation by shadowing effect of the inner disk parts. In these conditions, temperature in the planet-forming region can decrease down to very low values {($\sim$20 K; Mousis et al. 2009a)}. However, irradiation onto the central parts of the disk could modify the disk structure so much that shadowing is limited or absent in the outer parts. In this case, the temperature in the planet-forming region would be higher. On the other hand, the same model as the one used in this work has been used to explain the noble gas enrichments observed by the Galileo probe in the atmosphere of Jupiter (Owen et al. 1999; Mousis et al. 2009a). In order to account for the measured supersolar abundance of argon, the accretion of planetesimals formed at temperatures as low as $\sim$20 K has been invoked in the feeding zone of Jupiter. Here, because the K2V parent star of HD189733b is cooler than the Sun, we assume that the temperature and pressure conditions in the formation zone of HD189733b were as low as for Jupiter.

For each ice considered in Fig. 2, the domain of stability is the region located below its corresponding stability curve. The clathration process stops when no more crystalline water ice is available to trap the volatile species. Note that, in the pressure conditions of the disk, CO$_2$ crystallizes at a higher temperature than its associated clathrate. We then assume that solid CO$_2$ is the only existing condensed form of CO$_2$ in this environment. In addition, we have considered only the formation of pure ice of CH$_3$OH in our calculations because no experimental data concerning the equilibrium curve of its associated clathrate have been reported in the literature. The intersection of a thermodynamic path at given distance from the star with the stability curves of the different ices allows determination of the amount of volatiles that are condensed or trapped in clathrates at this location in the disk. Assuming that, once condensed, the ices add to the composition of planetesimals accreted by the growing planet along its migration pathway, this allows us to reproduce the volatile abundances by adjusting the mass of planetesimals that vaporized when entering the envelope. Our approach is supported by the simulations of Baraffe et al. (2006) that show that planetesimals are ablated in the envelope once the core mass reaches $\sim$6 \Mearth. Note that, because the migration path followed by the forming HD189733b is unknown, the 5--20 AU distance range of the ensemble of thermodynamic paths has been arbitrarily chosen to determine the composition of the accreted ices. The adoption of any other distance range for the planet's path beyond the snow line would not affect the composition of the ices (and thus HD189733b's global volatile enrichments) because it remains almost identical irrespective of i) their formation distance and ii) the input parameters of the disk, provided that the initial gas phase composition is homogeneous (Marboeuf et al. 2008). Table 2 gives the mean composition of ices incorporated in planetesimals formed {in the cold part} of the protoplanetary disk and ultimately accreted by proto-HD189733b. 

\begin{table}
\caption[]{Ratio of the mass of ice to the global mass of ices in planetesimals formed in the protoplanetary disk.}
\begin{center}
\begin{tabular}{lcc}
\hline
\hline
\noalign{\smallskip}
Ice		 		&  Mass fraction				\\	
\noalign{\smallskip}
\hline
\noalign{\smallskip}
H$_2$O			& $5.03 \times 10^{-1}$ 			\\
CO        			& $2.79 \times 10^{-1}$			\\
CO$_2$       		& $1.13 \times 10^{-1}$			\\
NH$_3$       		& $3.61 \times 10^{-2}$			\\
H$_2$S			& $2.37 \times 10^{-2}$			\\
N$_2$       		& $2.17 \times 10^{-2}$			\\
CH$_3$OH		& $1.94 \times 10^{-2}$			\\
CH$_4$   			& $3.44 \times 10^{-3}$    			\\
PH$_3$			& $8.24 \times 10^{-4}$			\\
\hline
\end{tabular}
\end{center}
\label{comp}
\end{table}
 
\section{Volatile enrichments inferred from the heavy element content}
 
Figure 1 shows the carbon and {oxygen} abundances (relative to solar) determined in HD189733b by S09 and MS09. Both studies suggest that carbon and oxygen abundances could be subsolar but MS09 also find that strongly supersolar elemental abundances are consistent with the spectra. If carbon only exists in the form of spectroscopically identified species, then the simultaneous fit of carbon and oxygen abundances retrieved by S09 in HD189733b requires the accretion of $\sim$1.2 \Mearth~of ices in the envelope (see Mousis et al. 2009b for details). This translates into the accretion of 2.3--4.6 \Mearth~of planetesimals in HD189733b's envelope if one assumes that the fraction of rocks and metals f$_{\rm r-m}$ varies between 0.47 and 0.74 in these solids, as for those formed in the outer Solar nebula (Johnson \& Lunine 2005; {Wong et al. 2008; Johnson \& Estrada 2009}). This mass range is well below the one predicted by interior models (20--80 \Mearth; Guillot 2008). Similarly, if each extreme abundance value found by MS09 is presumed representative of the composition of HD189733b's envelope, this also poses a problem of consistency with interior models. Indeed, considering the minimum carbon abundance of $\sim$1.2 $\times$ 10$^{-3}$ times the solar value inferred by MS09 in HD189733b, this corresponds to a maximum of $\sim$3 $\times$ $10^{-2}$ \Mearth~of heavy elements dissolved in the envelope, assuming a fraction of rocks and metals f$_{\rm r-m}$ of 0.74 in planetesimals. Alternatively, if one considers a carbon abundance of $\sim$35.6 times the solar value in HD189733b, i.e. the upper value retrieved by MS09, this directly translates into a minimum of $\sim$250 \Mearth~of heavy elements dissolved in the envelope, with f$_{\rm r-m}$ = 0 in planetesimals. We conclude that both extreme determinations found by MS09 require a mass of heavy elements that is well outside the 20--80 \Mearth~range predicted in HD189733b. These considerations imply that the determinations of S09 and MS09 are unable to constrain the mass range of heavy elements predicted by internal structure models.

\begin{figure}
\begin{center}
\resizebox{\hsize}{!}{\includegraphics[angle=-0]{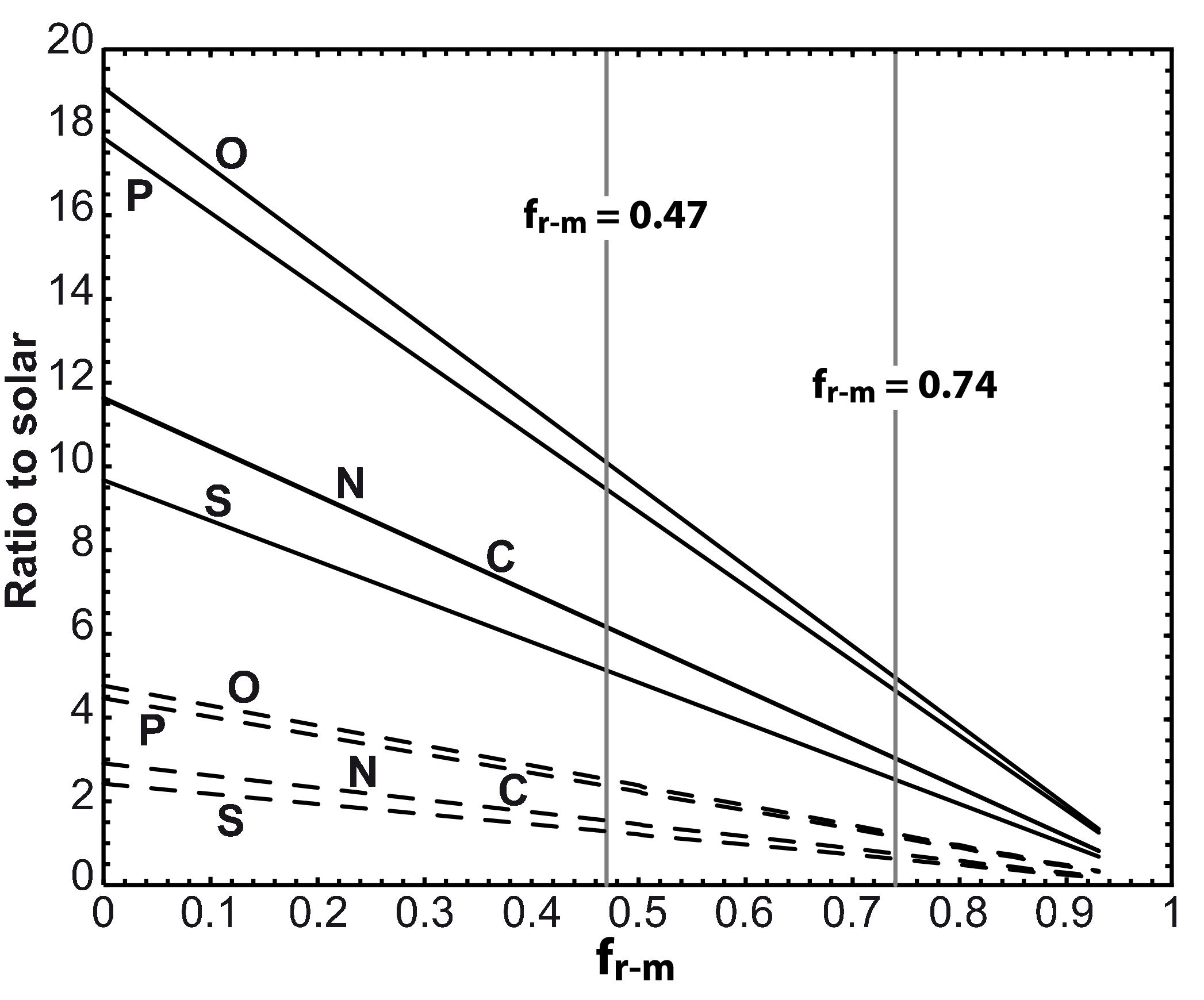}}
\caption{The volatile enrichments in the envelope of HD189733b as a function of the fraction of rocks and metals f$_{\rm r-m}$ in accreted planetesimals. Volatile enrichments have been calculated in the case of the presence of 20 \Mearth (dashed lines) and 80 \Mearth (solid lines) of heavy elements in the envelope, respectively. The two vertical lines enclose the range of plausible f$_{\rm r-m}$ values in planetesimals for which volatile enrichments are predicted in HD189733b (see text). N and C lines appear superimposed. {If 80 \Mearth~of heavy elements is present in the envelope, our enrichment predictions are within the range of determinations of MS09 from HST NICMOS spectrophotometry, irrespective of the f$_{\rm r-m}$ value in the 0.47--0.74 range. If 20 \Mearth~of heavy elements is present in the envelope, the simultaneous fit of the O and C determinations by MS09 from HST NICMOS spectrophotometry leads to a unique common solution for f$_{\rm r-m}$ = 0.48.}}
\end{center}
\label{heavy}
\end{figure}

Figure 3 represents the volatile enrichments (relative to solar) in the envelope of HD189733b as a function of the fraction of rocks and metals in accreted planetesimals. Note that the enrichment values of S, N and P are shown together with those of O and C because they may be tested by future observations of HD 189733b's atmosphere. {In particular, S is expected to be contained mainly in H$_2$S at altitudes of 0.002--1 bar in HD 189733b (Zahnle et al. 2009). Its abundance at these levels may be sampled provided observations at sufficiently high spectral resolution of its strong 2 $\mu$m feature become available. The elemental abundances of N and P are likely more difficult to measure in HD 189733b. The main nitrogen-bearing and phosphorus-bearing gases are predicted to be N$_2$ and P$_2$ (Visscher et al. 2006). Both species lack a dipole moment, and are therefore challenging to measure at IR wavelengths. In exoplanets cooler than HD 189733b, elemental N and P are mainly in NH$_3$ and PH$_3$ forms below altitudes where photolysis occurs (Saumon et al. 2000; Visscher et al. 2006), and these molecular species are spectroscopically much easier to detect.} The volatile enrichments  have been calculated for the two extreme values covering the 20--80 \Mearth~range of heavy elements estimated to be present in HD189733b (Guillot 2008). If 20 \Mearth~of heavy elements are dissolved in the envelope, O, C, N, S and P are found to be 1.2--2.5, 0.8--1.5, 0.8--1.5, 0.6--1.3, and 1.2--2.3 times supersolar for f$_{\rm r-m}$ ranging between 0.47 and 0.74 in planetesimals, respectively. In contrast, if 80 \Mearth~of heavy elements are present in the envelope, then O, C, N, S and P become 5.0--10.1, 3.0--6.2, 3.0--6.2, 2.5--5.1, and 4.6--9.5 times supersolar for the same f$_{\rm r-m}$ range in planetesimals, respectively. The comparison of our calculations with the full range of subsolar C and O abundances retrieved by MS09 from Spitzer broadband photometry data suggest that their values do not correspond to the global heavy element enhancement presumed to exist in HD189733b. In contrast, when comparing our enrichment predictions with the supersolar values retrieved by MS09 from HST NICMOS spectrophotometry, we find that our values calculated for a mass of 80 \Mearth~of heavy elements present in the envelope are within their range of determinations, irrespective of the f$_{\rm r-m}$ value. In the case of 20 \Mearth~of heavy elements present in the envelope, the simultaneous fit of the O and C determinations by MS09 from HST NICMOS spectrophotometry leads to a unique common solution, which corresponds to O, C, N, S and P found to be 2.5, 1.5, 1.5, 1.3, and 2.3 times supersolar for f$_{\rm r-m}$ = 0.48, respectively.

\section{Influence of hidden carbon}

\begin{figure}
\begin{center}
\resizebox{\hsize}{!}{\includegraphics[angle=-0]{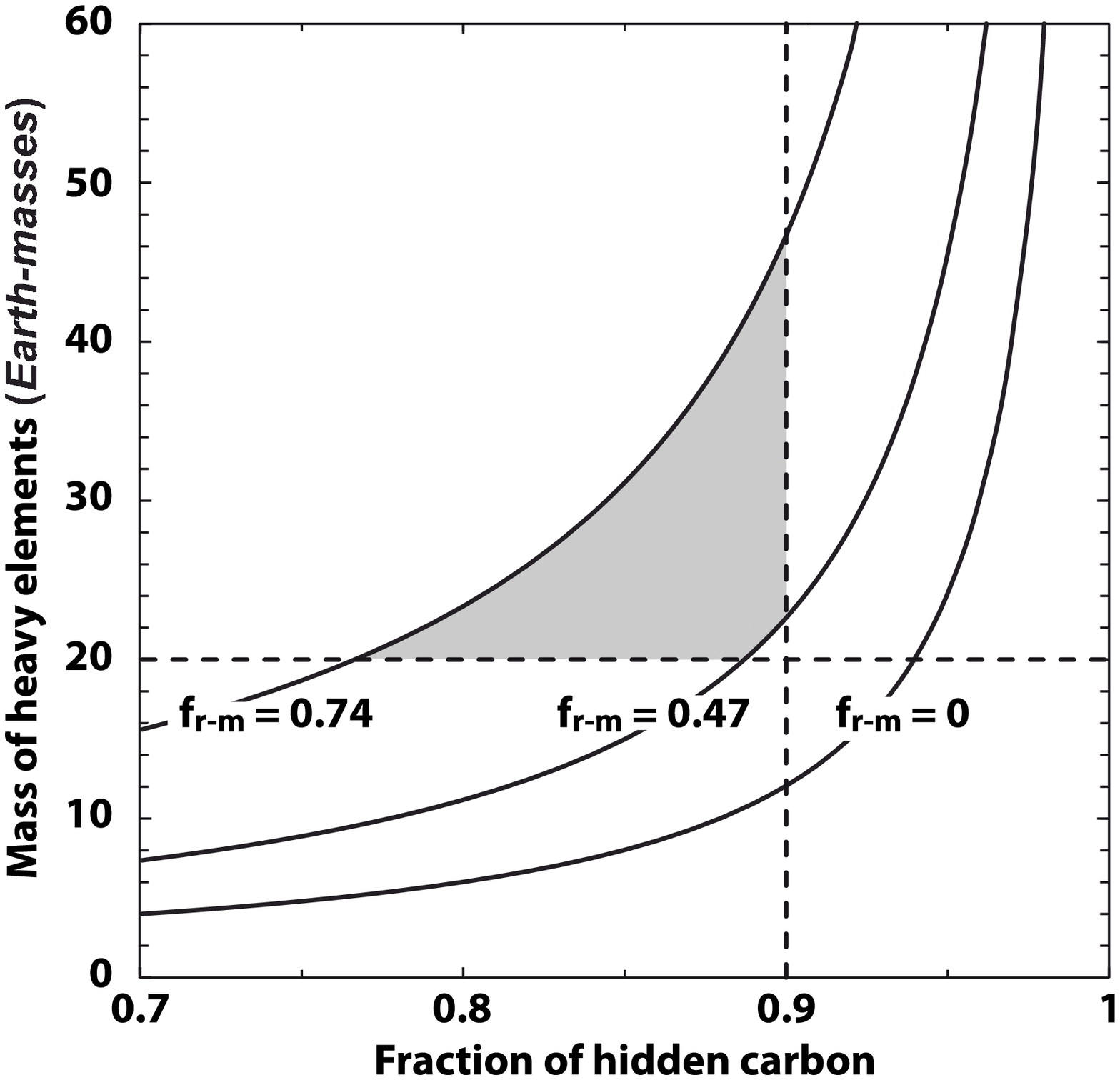}}
\caption{The mass of heavy elements in HD189733b as a function of the fraction of carbon hidden in the atmosphere (see text). The horizontal dashed line corresponds to the minimum mass of heavy elements predicted in HD189733b (Guillot 2008). The vertical dashed line corresponds to the upper value of the carbon fraction that can exist in the form of soots and organic molecules at sampled pressure levels. The three solid curves correspond to different fractions of rocks and metals f$_{\rm r-m}$ in planetesimals accreted by HD189733b during its formation (see text). The grey area represents the range of fractions of hidden carbon matching the global heavy element content predicted in HD189733b's interior, which is enclosed by the two most plausible values of f$_{\rm r-m}$, assuming a composition of planetesimals similar to those formed in the Solar nebula (see text).}
\end{center}
\label{heavy2}
\end{figure}

Recent photochemical models of HD189733b's atmosphere suggest that a substantial part of carbon could be present in the form of etylene, acetylene, PAHs and probably soots at the sampled pressure levels, implying that the retrieved carbon abundances could be under-estimated (Marley et al. 2009; Zahnle et al. 2010). Depending on the efficiency of the vertical mixing and for an optimum temperature of $\sim$1000 K in the envelope, up to $\sim$90\% of carbon could exist in the form of soots and organic molecules (Zahnle et al. 2010). In these conditions, a range of the subsolar abundances inferred in the planet could be consistent with its metal content if a fraction of carbon is postulated to exist in the form of the not yet detected molecules. Figure 4 represents the mass of heavy elements in HD189733b as a function of the fraction of ``hidden'' carbon at the sampled pressure levels, assuming that the abundances of spectroscopically visible oxygen and carbon are within the range of values found by S09. In this case, the observable oxygen and carbon abundances have been set to $\sim$0.3 and 0.15 times solar, i.e. values corresponding to the fit of our model to the highest oxygen abundance found by S09. The figure shows that, if 77 to 90\% of carbon is contained in these compounds, then the mass of heavy elements needed to fit the observed volatile abundances is now between $\sim$20 and 46.5 \Mearth~and matches the mass range predicted by interior models for the 0.47--0.74 range of f$_{\rm r-m}$ values in planetesimals. In this case, our model also predicts an oxygen abundance ranging between 1.3 and 2.9 times the solar value in HD189733b's envelope. 

\section{Alternative scenarios}

{Several alternative hypotheses to the presence of soots} in HD189733b can be formulated in order to account for the possible discrepancy between the elemental atmospheric abundances and the global budget of heavy elements in the planet. This discrepancy could have been engendered by processes that, similarly to the hypothesis of the presence of soots previously detailed, take place in the atmosphere of HD189733b during its postaccretion evolution. For example, a plausible alternative could be the presence of {\it differential settling} in HD189733b, which would result from the combination of gravity and irradiation effects that took place inside the atmosphere of HD189733b, thus lowering the C and O (and any other volatile) abundances in the upper layers (Mousis et al. 2009b). In this case, the measured C and O atmospheric abundances would not be representative of the envelope composition because the atmosphere is isolated from the interior by a radiative zone. This idea has not yet been investigated by numerical modeling.

On the other hand, the assumptions of our model depicting the volatile enrichments in HD189733b are based on a given mass range of accreted solids defined by interior models (Guillot 2008), on the temperature at which the solids/ices condensed in the solar nebula, and on the presence of a core with a defined size.  The basis of our comparison here is Jupiter, but {\it HD 189733b may not have followed the same accretional and dynamical history as Jupiter did} because it migrated so far inward. For example, our volatile abundance calculations have been performed assuming that the core of HD189733b has eroded with time, implying that all the heavy elements are included in the envelope. This assumption derives from Jupiter's interior models that also predict that the mass of its core may actually be as low as zero (Saumon \& Guillot 2004). Neglecting the possibility of erosion and adopting a core mass corresponding to the maximum one predicted for Jupiter ($\sim$13 \Mearth; Saumon \& Guillot 2004) would decrease our volatile enrichment predictions by a factor between $\sim$0.35 and 0.84, depending on the considered total mass of heavy elements (20 to 80 \Mearth) in HD189733b. In the case of a 0.35 factor, the O and C predicted abundances already become 0.4 and 0.3 times solar, respectively. If the core mass of HD189733b is equal to 15 \Mearth, and assuming a total mass of heavy elements of 20 \Mearth~in the planet, then the O and C predicted abundances become respectively 0.3 and 0.2 times solar and are within the range of subsolar abundances retrieved by MS09. The assumption of any larger core in HD189733b would imply lower abundances of heavy elements in the envelope. 

An alternative explanation is that the mass of heavy elements present in HD189733b is lower than the 20--80 \Mearth~mass range in the models preferred by Guillot (2008) on the basis that the same hypothesis (modified equation of state or increased opacity scenario) allows to account for the properties of all known transiting giant extrasolar planets. {Indeed, while the 20-80 \Mearth~mass range of heavy elements is based on an explanation consistent for the masses-radii of all  planets, the mass and radius of HD189733b itself is consistent with a larger mass range (including zero).} A low mass of heavy elements in HD189733b could be due to the formation of large planetesimals by the time the giant planet accreted its gas envelope. Planetesimals would then dynamically decouple from and not necessarily be accreted with the gas, and thus the giant planet's gas envelope would actually be metal-poor. This would leave the planet with a heavy-element poor envelope if core erosion did not take place and little planetesimal accretion occurred after the giant planet grew to Jupiter size. The key point is whether planetesimal mass is dynamically coupled to the gas or not. If most of the mass of solids is in lunar-size objects, then this mass will certainly not be coupled to the gas. Another question is related to what extent C and O condense out of the gas because if these volatiles were not incorporated in planetesimals, then they should remain in solar abundances in the planet's envelope.
 
Another possibility is that the subsolar abundances in HD189733b would result from the {\it accretion of volatile-poor planetesimals}, if one postulates that the heavy element budget of the planet was mainly acquired inward of the nebula snow line. This might run into difficulty because of the eventual poor efficiency of the planetesimals accretion during the {formation} of HD189733b: dynamical models of Guillot \& Gladman (2000) suggest that the ratio of accreted to ejected planetesimals converges towards zero in the case of giant planets with masses reaching or exceeding that of Jupiter. However, the simulations of {Guillot \& Gladman (2000) were performed without addressing the inwards migration of giant planets}. In this case, as planets move inwards, the Safronov number decreases and more accretion could occur. This scenario can lead to two extreme possibilities. The first possibility consists in the accretion of gas and planetesimals at the same location and at temperatures greater than $\sim$160 K in the disk (i.e. the condensation temperature of water) by the growing HD189733b. In this case, the envelope of the planet should contain solar abundances of volatiles because the accreted gas would have never been fractionated by their condensation. In the second possibility, if the planet accreted most of its gas at very low temperature ($\sim$20 K) beyond the snowline, migrated inwards and accreted its heavy elements in the warm part of the disk, then it should display oxygen and carbon subsolar abundances. Indeed, the gas accreted by the planet should be oxygen and carbon {poor} because these species would have {mostly} formed pure ices or been incorporated into clathrates that decoupled from gas. An intermediate case could be that the temperature of the gas accreted by the planet beyond the snow line is low enough to allow the condensation of water ($\le$ 160 K) but remains too high to allow the condensation or trapping of carbon ($>$ 100 K). In these conditions, the oxygen abundance should be subsolar in the envelope whilst that of carbon remains solar.

Finally, other alternative possibilities can be envisaged but none of them is found satisfying. Indeed, the subsolar C and O abundances in HD189733b could result from the fact that the {\it planetesimals accreted by the growing planet did not ablate in the atmosphere} but carried their volatiles intact  to the deep interior, where they remain. However, simulations by Baraffe et al. (2006) show that 100 km planetesimals are destroyed in the envelope once the core mass reaches around 6 \Mearth, making this hypothesis unlikely. Alternatively, it is possible for {\it oxygen to be depleted relative to carbon inward of the snow line} thanks to the cold-trapping or cold-finger effect of water forming ice at the snowline and drying the nebula inward, but this event involves principally oxygen  (Stevenson \& Lunine 1988; Cyr et al. 1999) and does not account for the subsolar abundance of carbon.

\section{Conclusions}

Using a model describing the formation sequence and composition of planetesimals in the protoplanetary disk, we determined the range of volatile abundances in the envelope of HD189733b that is consistent with the 20--80 \Mearth~of heavy elements estimated to be present in the planet's envelope. This model has been used in the framework of the core accretion model but the results obtained here may be valid in the case of the gravitational instability model if planetesimals accretion was possible during or after the collapse of gas to form HD189733b. Assuming that carbon exists only in the form of spectroscopically detected species, we find that none of the volatile abundances determined from our model is consistent with the volatile subsolar abundances retrieved by S09 and MS09. The same statement applies to the largest supersolar volatile abundances inferred by MS09 from which the corresponding mass of heavy elements largely exceeds the mass range derived from interior models. In contrast the volatile abundances inferred from the mass range of heavy elements estimated to be present in the planet are found within the range of supersolar abundances determined by MS09 from HST NICMOS photometry.

On the other hand, the presence of soots, PAHs and organic molecules as suggested by the photochemical model of Zahnle et al. (2010) in HD189733b's envelope could compensate for the subsolar carbon abundances inferred from spectroscopically active species.  The presence of the additional carbon-bearing species leads to supersolar abundances that match the predicted heavy element content for this planet, as was the case in an earlier study of Jupiter (Mousis et al. 2009a). The calculated oxygen abundance in HD189733b is larger than that inferred from observations but the discrepancy could have resulted from an oxygen depletion in the envelope during its thermal history, due to the immiscibility of oxygen and hydrogen under high temperature conditions (Fortney \& Hubbard 2003), again similar to a possible evolutionary scenario for Jupiter and Saturn that also accounts for their apparent oxygen deficiencies (Owen et al. 1999; Gautier et al. 2001; Mousis et al. 2009a). A preliminary diagnostic test that would strongly support the presence of soots in the atmosphere of HD189733b is the detection of molecular precursors such as acetylene (C$_2$H$_2$) via the measurement of its C-H stretch at 3.03 $\mu$m in the brightest part of the infrared spectrum. This band has been fully characterized in the laboratory up to 1500 K (Amyay et al. 2009). At the temperature of $\sim$1000 K, C$_2$H$_2$ initiates a carbon-rich chemistry that ultimately leads to the formation of soots (Zahnle et al. 2010) via a large number of different potential physical and chemical pathways. The measurement of the aromatic/aliphatic content of the carbonaceous material embedded in the envelope of HD189733b, which allows one to trace its stage of evolution, could be constrained via the measurement of the C-H stretch absorption features at 3.3 $\mu$m (aromatic compounds) and 3.4 $\mu$m (aliphatic compounds) (Dartois et al. 2004). Similarly, the position of the C-C aromatic stretch region can also be used as a tracer of the aromaticity of the material at 6.2 $\mu$m (aromatic compounds) and 6.3 $\mu$m (aliphatic compounds). Because it is difficult to investigate directly the spectral features of soots  (J{\"a}ger et al. 2009; Pino et al. 2008; Biennier et al 2009), spectroscopic identification of chemical precursors is the best way to test the model described in this paper.

{Several alternative hypotheses} to the presence of soots in HD189733b have also been formulated in order to reconcile the apparent discrepancy between the observed elemental atmospheric abundances and the global budget of heavy elements in the planet inferred by interior models. Among these hypotheses, the possibility of differential settling in the envelope of HD189733b, the presence of a larger core that did not erode with time, a mass of heavy elements lower than the one predicted by interior models, a heavy element budget resulting from the accretion of volatile-poor planetesimals in specific circumstances or the combination of all the mechanisms invoked in this work could also explain the observed subsolar elemental abundances.

\acknowledgments
We thank J.-P. Beaulieu and G. Tinetti for their valuable comments on the manuscript. OM acknowledges support from CNES. JIL's work was supported within the scope of the program `Incentivazione alla mobilita' di studiosi straineri e italiani residenti all'estero. We acknowledge T. Guillot for enlightning discussions and information on his work. We thank an anonymous reviewer for his constructive comments which helped us improve our manuscript.

%

\end{document}